\pdfoutput=1 
\documentclass[aps,amsmath,amssymb,reprint]{revtex4-1}
\usepackage{graphicx}
\usepackage{subcaption}
\usepackage{bm}
\usepackage{amsmath}
\usepackage{dcolumn}
\usepackage{amssymb}
\usepackage{mathtools}
\usepackage[colorlinks,pdfstartview=FitH]{hyperref}
\hypersetup{linkcolor=blue,citecolor=blue,filecolor=black,urlcolor=blue}

\newcommand\be{\begin{eqnarray}}
\newcommand\ee{\end{eqnarray}}

\begin{document}

\title[The Chiral Qubit]{The Chiral Qubit:\\quantum computing with chiral anomaly}

\author{Dmitri E. Kharzeev}
\email{dmitri.kharzeev@stonybrook.edu.}
\affiliation{Department of Physics and Astronomy, Stony Brook University,\\ New York 11794-3800, USA}

\affiliation{Department of Physics and RIKEN-BNL Research Center,\\ Brookhaven National Laboratory,  Upton, New York 11973, USA}
\
\author{Qiang Li}
 \email{qiangli@bnl.gov.}
\affiliation{Condensed Matter Physics and Materials Science Department,\\ Brookhaven National Laboratory, Upton, NY 11973, USA}

\date{\today}

\begin{abstract}
The quantum chiral anomaly enables a nearly dissipationless current in the presence of chirality imbalance and magnetic field -- this is the Chiral Magnetic Effect (CME), observed recently in Dirac and Weyl semimetals. Here we propose to utilize the CME for the design of qubits potentially capable of operating at THz frequency, room temperature, and the coherence time to gate time ratio of about $10^4$. The proposed "Chiral Qubit" is a micron-scale ring made of a Weyl or Dirac semimetal, with 
the $|0\rangle$ and  $|1\rangle$ quantum states corresponding to the symmetric and antisymmetric superpositions of quantum states describing chiral fermions circulating along the ring clockwise and counter-clockwise. 
A fractional magnetic flux through the ring induces a quantum superposition of the $|0\rangle$ and  $|1\rangle$ quantum states. 
The entanglement of qubits can be implemented through the near-field THz frequency electromagnetic fields.
\end{abstract}

\maketitle

\section{Introduction}
Quantum computing based on superconducting qubits has made an impressive progress recently, see \cite{gambetta2017building} for a review. However, the reliance on superconductivity imposes severe constraints on quantum processors: one needs to cool the qubits to low ($\sim$ 10 mK) temperatures, and the magnitude of superconducting gap limits the operation frequency of quantum processors to the GHz range. Therefore, making a leap towards the mass-producible and faster quantum computers requires new ideas. In this paper we propose to replace the superconducting qubits by the ``chiral qubits" utilizing the Chiral Magnetic Effect predicted in the context of high energy physics \cite{Kharzeev2008,Kharzeev2009} (see \cite{Kharzeev2014,kharzeev2013strongly} for reviews and additional references) and discovered recently \cite{Li2016,Xiong2015,Huang2015} in 3D chiral materials.
\vskip0.3cm

The Chiral Magnetic Effect (CME)  is the generation of electric current induced by chirality imbalance in the presence of magnetic field. 
It is a macroscopic manifestation of the chiral anomaly \cite{Adler1969,Bell1969} in relativistic field theory of chiral fermions (massless spin $1/2$ particles with a definite projection of spin on momentum) -- a quantum phenomenon arising from a collective motion of particles and antiparticles in the Dirac sea perturbed by the gauge fields. The advent of Dirac and Weyl semimetals \cite{Wang2012,Wang2013,Borisenko2014,Liu2014} (see \cite{armitage2018weyl} for a review) with chiral quasi-particles opened the possibility to study the effects of chiral anomaly in condensed matter \cite{Nielsen1983}.
The chiral anomaly creates chirality imbalance in parallel electric and magnetic fields and thus enables CME~\cite{Kharzeev2008,Kharzeev2009}. The resulting longitudinal negative magnetoresistance~\cite{Son2013, burkov2015negative} has been observed in Dirac semi-metals such as $\mathrm{ZrTe_5}$~\cite{Li2016} and $\mathrm{Na_3 Bi}$~\cite{Xiong2015} and Weyl semi-metals such as TaAs~\cite{Huang2015}.
\vskip0.3cm

In this paper we propose a design of a "chiral qubit" based on this novel quantum phenomenon.
The advantages of the proposed qubit architecture stem from the presence of CME at much higher temperatures $\sim$ 150 K (and potentially at room temperature), and the predicted \cite{kaushik2019chiral} possibility to manipulate the chiral magnetic current by light with $\sim$ 10 THz frequency. At the same time, it appears that the Hamiltonian describing the chiral qubit is very similar to the Hamitonian of superconducting qubits, which enables the traditional implementation of quantum gates. While the chirality coherence time in presently available chiral materials is quite short, with the chirality flipping rate in the GHz frequency range, the ability to operate quantum gates with  $\sim$ 10 THz frequency leads to the ratio of coherence time to gate time of about $10^4$ that is sufficient for the implementation of quantum error correction algorithms \cite{lidar2013quantum}.
\vskip0.3cm

The paper is organized as follows. In section \ref{sec:anomaly} we briefly summarize the known features of chiral anomaly that enable the chiral qubit, and adapt this discussion to the geometry of the proposed device -- a ring made of a Dirac or Weyl semimetal. In section \ref{sec:qubit}, we describe the chiral qubit, discuss its Hamiltonian, and show that it is very similar to the Hamiltonian describing  superconducting qubits. Finally, in section \ref{sec:outlook} we point out the advantages and disadvantages of the proposed chiral qubit, and discuss the possible path towards realization of the proposed device.

\section{\label{sec:anomaly}Chiral anomaly}

Dirac and Weyl semimetals possess quasiparticles that at low energies behave as massless relativistic fermions. Let us thus begin with a brief review of relativistic fermions in $(3+1)$ and  $(1+1)$ dimensions interacting with electromagnetic fields, and introduce the chiral anomaly \cite{Adler1969,Bell1969}; we will adapt the discussion to chiral materials.
\vskip0.3cm

Charged massless fermions propagating with velocity $v_F$  are described by the Dirac Lagrangian
\be\label{lagr}
{\cal L} = \bar\psi \{i \gamma^0 D_0 + i v_F \gamma^i D_i\} \psi - \frac{1}{4} F_{\mu\nu}^2,
\ee   
where $D_\mu = \partial_\mu + i e A_\mu$, $\psi$ is the Dirac spinor that has 4 components in $(3+1)$ dimensions, and 2 components in $(1+1)$ dimensions, $\gamma_\mu$ is the set of Dirac matrices satisfying the anticommutation relations $\{\gamma_\mu, \gamma_\nu\} = 2 g_{\mu\nu}$, $\bar\psi \equiv \psi^\dagger \gamma^0$; the Minkowski metric is defined as $g^{\mu\nu} = \rm{diag} (+1,-1,-1,-1)$ and $g^{\mu\nu} = \rm{diag} (+1,-1)$ in $(3+1)$ and $(1+1)$ dimensions respectively; we set the Planck constant $\hbar =1$ for now, and will restore it when needed.

The peculiarity of the $(1+1)$ dimensional fermion theory that we will use is that the gauge potential $A_{\mu} = A_{\mu}(\vec x, t)$ and the corresponding field strength tensor $F_{\mu\nu}$ will always ``live" in $(3+1)$ dimensions -- this corresponds to a thin wire or ring of a chiral material embedded into a physical $(3+1)$ dimensional space, and is different from the massless $(1+1)$ QED known as the Schwinger model \cite{schwinger1962gauge}. Nevertheless some features of the Schwinger model \cite{coleman1975charge}, especially those dictated by the chiral anomaly \cite{manton1985schwinger}, emerge in our case as well.
\vskip0.3cm

Massless fermions possess a conserved quantum number -- chirality. For positive energy states (particles), chirality is equal to helicity, i.e. the projection of spin on momentum in $(3+1)$ dimensions, and direction of momentum in  $(1+1)$ dimensions. For negative energy states (holes), chirality is equal to {\it minus} helicity.

\begin{figure}[htbp]
  \begin{subfigure}{9cm}
\hspace{-1.3cm}\centering\includegraphics[width=9cm]{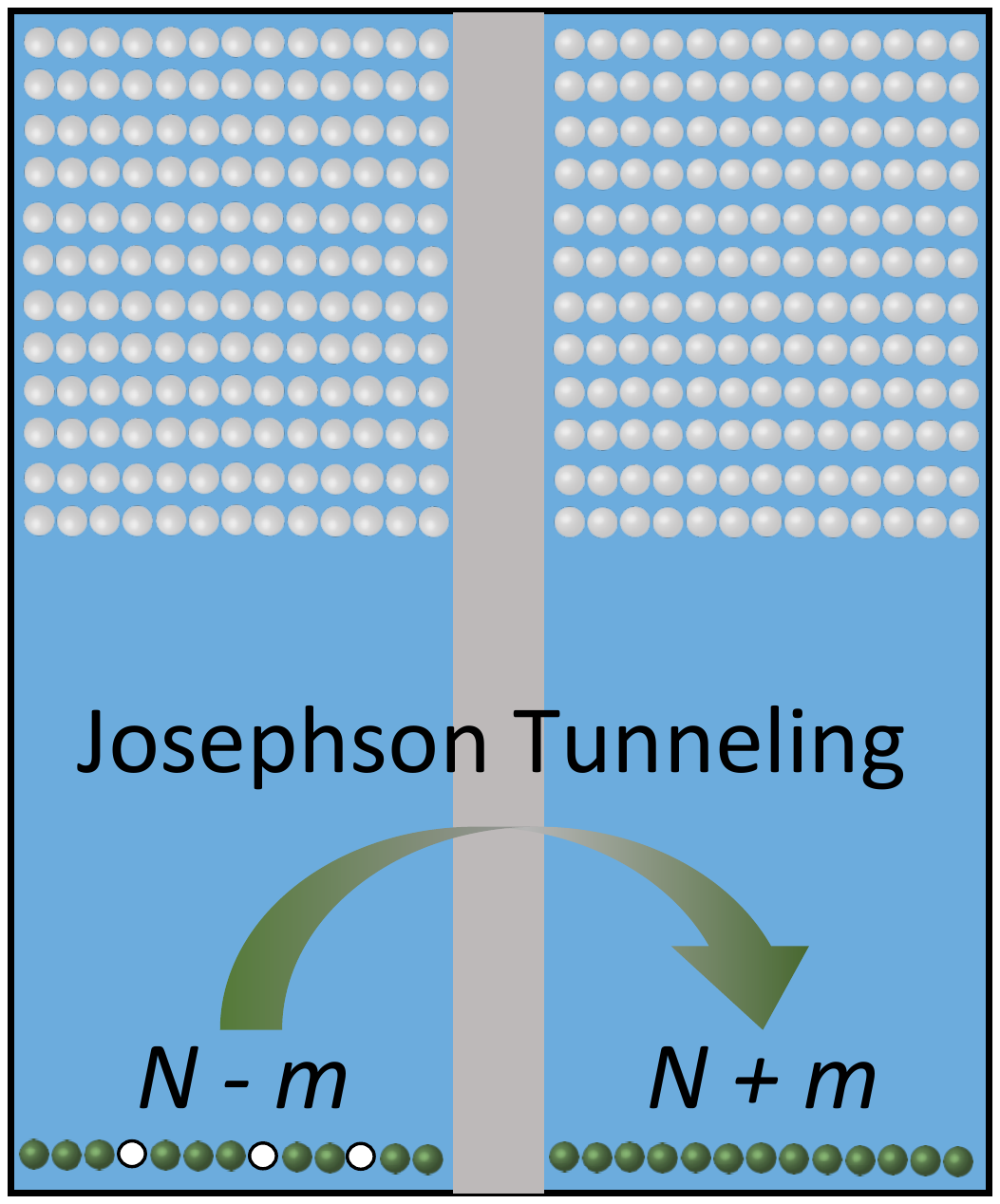}
    \vspace{-1cm}
    \caption{Josephson tunneling between the two Cooper boxes separated by an insulator or a normal metal.}
  \end{subfigure}
  \begin{subfigure}{8cm}
    \hspace{-2.5cm}\centering\includegraphics[width=8cm]{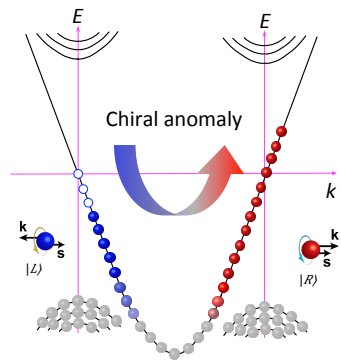}
    \caption{Chiral charge pumping between the left- and right-handed cones in a Weyl semimetal.}
  \end{subfigure}
  \caption{Comparison of the Josephson tunneling used in a superconducting transmon qubit (a) and the chiral charge pumping used in the proposed chiral qubit (b).}
  \label{imbalance}
\end{figure}

It is convenient to introduce the spinors $\psi_L = P_L \psi$ and $\psi_R = P_R \psi$ corresponding to the fermions of left and right chirality by using the projection operator $P_{R/L} = \frac{1}{2}(\hat{\rm I} \pm \gamma_5)$, with $\gamma_5 \equiv  i \gamma_0 \gamma_1 \gamma_2 \gamma_3$ in $(3+1)$ dimensions and $\gamma_5 \equiv \gamma_0 \gamma_1$ in $(1+1)$ dimensions \footnote{In even number of space-time dimensions $d=s+1$ (odd number of spatial dimensions $s$) the generalization of $\gamma^5$ matrix is defined as $\gamma^{d+1} = \sqrt{(-1)^{(1-s)/2}} \gamma^0 ... \gamma^d$.}; $\hat{\rm I}$ is a $4\times4$ or $2\times2$ unity matrix in the corresponding number of dimensions. 

In terms of left- and right-handed chiral fields, the vector (electric) current carried by Dirac fermions is given by $J^\mu = \bar\psi \gamma^\mu \psi = J^\mu_R + J^\mu_L$, where $J^\mu_{R,L} = \psi_{R,L} \gamma^\mu \psi_{R,L}$. The current of chirality, or the axial current, is defined as $J^\mu_A = \bar\psi \gamma^\mu \gamma^5 \psi$, and can be decomposed as $J^\mu_A = J^\mu_R - J^\mu_L$. It is easy to see that under the parity transformation $\cal P$ that interchanges $\psi_{R/L}(\vec x, t) \to \psi_{L/R}(- \vec x, t)$, the spatial component of the vector current is odd: $J^i \to - J^i$, and that of the axial current is even, $J_A^i \to J_A^i$. 

The conservation of chirality  follows from the global $U_A(1)$ symmetry of the Lagrangian  (\ref{lagr}) 
describing the massless limit of Dirac theory. Indeed, by using Dirac equation, one finds that the divergence of the axial current generated by the $U_A(1)$ transformation $\psi(x) \to \exp(i \alpha \gamma_5) \psi(x)$ is given by
\be
\partial_\mu J^{\mu}_A = 2 i m \bar{\psi} \gamma_5 \psi ,
\ee
so that the axial current is conserved in the massless $m \to 0$ limit.

\vskip0.3cm

It appears however that quantum effects arising from the interactions of charged fermions with electromagnetic field spoil the conservation of the axial current $J^{\mu}_A$, so the corresponding chiral charge $Q_A = \int d^3x J^0_A$ is no longer conserved. The chiral anomaly appears in any even number of space-time dimensions. In $(3+1)$ dimensions, in the chiral limit of massless fermions one gets \cite{Adler1969,Bell1969} 
\be
\partial_\mu J^{\mu}_A = \frac{e^2}{16 \pi^2}\ \epsilon^{\mu\nu\rho\sigma} F_{\mu\nu} F_{\rho\sigma} ,
\ee
where the quantity on the r.h.s. is proportional to the scalar product of electric and magnetic fields. 
In the case of $(1+1)$ dimensions, the chiral anomaly is driven by an external electric field:
 \be\label{anomaly_2}
\partial_\mu J^{\mu}_A = \frac{e}{2 \pi}\ \epsilon^{\mu\nu} F_{\mu\nu} .
\ee
Let us discuss the physical meaning of the chiral anomaly (\ref{anomaly_2}). In $(1+1)$ dimensions
we deal with two-component spinors, and can satisfy the Dirac algebra by using the Pauli matrices; we choose $\gamma_0 = \sigma_x$, 
and $\gamma_1 = - i \sigma_y$. We thus have $\gamma_5 \equiv \gamma_0 \gamma_1 = \sigma_z$. We can now use the eigenstates $\psi_R$ and $\psi_L$ of $\gamma_5 = \sigma_z$ corresponding to eigenvalues $+1$ and $-1$ respectively to represent the spinor $\psi$:
\be
\psi = \begin{pmatrix} \psi_R \\ \psi_L \end{pmatrix} .
\ee
This is completely analogous to the decomposition of spinor in nonrelativistic quantum mechanics in terms of eigenstates of $\sigma_z$ that represent the spin-up and spin-down configurations. 
\vskip0.3cm

Consider now the Lagrangian (\ref{lagr}) in  $(1+1)$ dimensions; let us write down the fermion part in terms of fields $\psi$ and $\psi^\dagger$ as (recall that $\bar\psi \equiv \psi^\dagger \gamma^0$, and $\gamma_0^2 = 1$)
\be\label{lagr_ferm}
\hspace{-0.4cm} {\cal L} = \psi_R^\dagger \{i (D_0 + v_F D_1) \} \psi_R  + \psi_L^\dagger \{i (D_0 - v_F D_1)\} \psi_L ,
\ee
where we have introduced the covariant derivatives $D_\mu \equiv \partial_\mu + i e A_\mu$.
It is easy to see that the chiral projector $P_{R/L} = \frac{1}{2}(\hat{\rm I} \pm \gamma_5) = \frac{1}{2} (\hat{\rm I} \pm \sigma_z) $ selects the left- and right-moving particles described by $\psi_R = P_R \psi$ and $\psi_L = P_L \psi$ propagating along the left and right edges of the ``light cone" defined by $p_0^2 - v_F^2 |p|^2 =0$. Note that the massless lagrangian (\ref{lagr_ferm}) does not mix the right- and left-handed fermions. 
\vskip0.3cm

We are now ready to examine the meaning of the chiral anomaly (\ref{anomaly_2}). The r.h.s. of  (\ref{anomaly_2}) is proportional to a full derivative: $\epsilon^{\mu\nu} F_{\mu\nu} = 2 \partial_\mu \left(\epsilon^{\mu\nu} A_\nu \right)$. Using the Coulomb gauge $\partial_1 A_1 = 0$ in which the electric field is $E_1 = - \partial_0{A}_1$ and $\epsilon^{\mu\nu} F_{\mu\nu} = -2 E_1$, we see that this full derivative term is relevant in  the presence of  a non-zero electric field:
\be\label{anom_e}
\partial_\mu J^{\mu}_A = - \frac{e}{\pi}\ E_1 .
\ee 
Consider the effect of a time-dependent electric field on the fermions. The total chiral charge of the system carried by the fermions is given by the difference in the number of the right- and left-handed fermions: $Q_A = N_R - N_L = \int dx\ J^0_A$. By integrating (\ref{anom_e}) over the (one-dimensional)   space, we get
\be\label{anom_eq}
\frac{d(N_R-N_L)}{dt} =  - \frac{e}{\pi}\ \int dx E_1 = \frac{e}{\pi}\ \frac{d}{dt}\int dx A_1.
\ee 
This relation shows that chirality can be transferred between the fermions and the gauge field, and only the sum of fermion chirality and the gauge field chirality expressed through the Chern-Simons 1-form $\int dx A_1$ is conserved. An external electromagnetic field can create a chiral imbalance between the number of the right- and left-handed fermions as shown in Fig. \ref{imbalance} b. The sign of the imbalance is determined by the sign of Chern-Simons 1-form $\int dx A_1$, or equivalently by the phase of the Wilson line $W[A] \equiv \exp\{i e  \int dx A_1\}$. 
\begin{figure}[htbp]
\includegraphics[width=7cm]{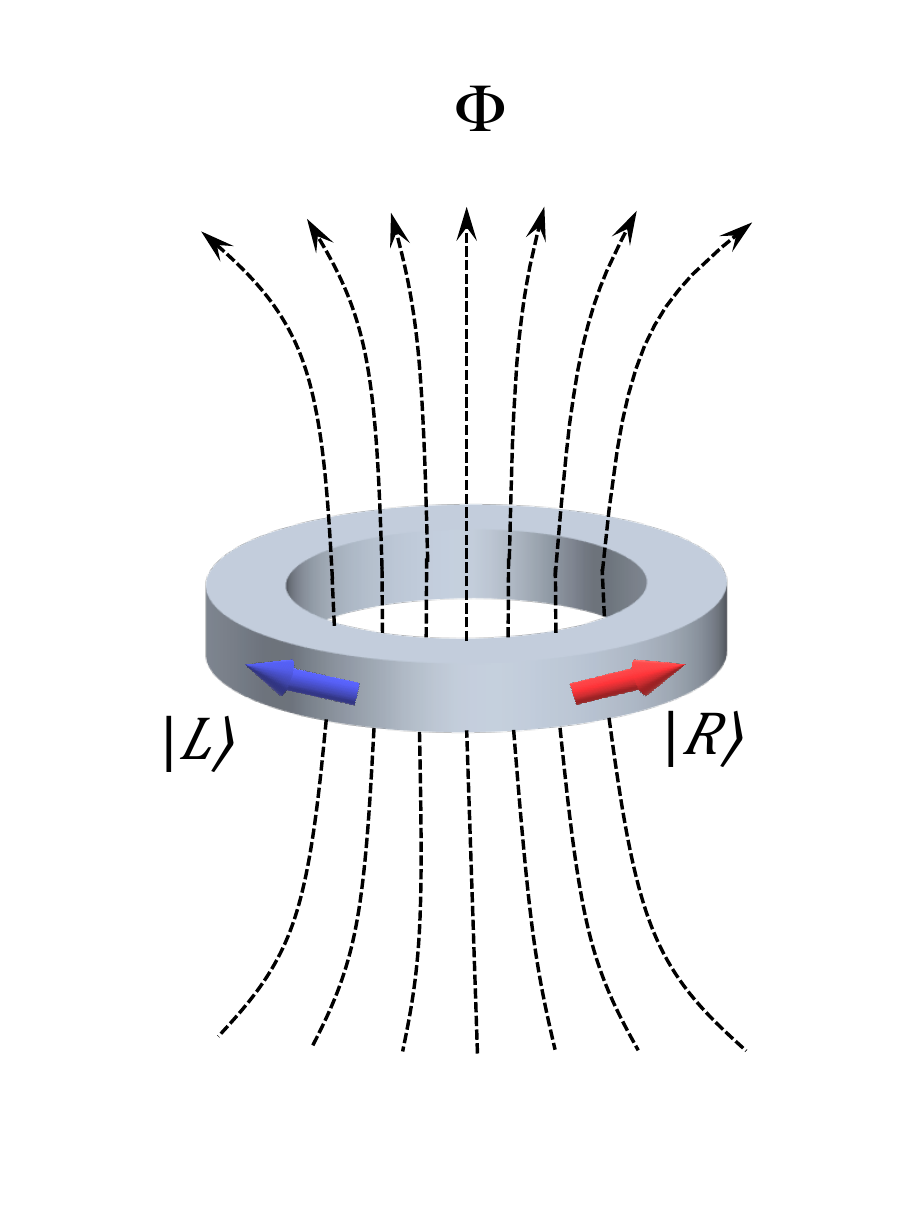}
\caption{The chiral qubit: a ring made of a Dirac or Weyl semimetal. The clockwise and counter-clockwise quantum currents form the $|R\rangle$ and $|L\rangle$ states controlled by an external electromagnetic field. A fractional flux through the ring creates a quantum superposition of the $|R\rangle$ and $|L\rangle$ states.}
\label{chiral_qubit}
\end{figure}
%
\vskip0.3cm
The dynamics of chiral anomaly as described above is very similar to the dynamics of superconducting ring in an external magnetic field. It is also similar to the tunneling of Cooper pairs through the Josephson junction in a charge qubit. Indeed, consider a charge (or transmon) qubit shown schematically in Fig. \ref{imbalance} a) that consists of two superconducting islands (or ``Cooper boxes") separated by an insulator; for a concise review, see \cite{girvin2011circuit}. The Cooper pairs can tunnel through the junction creating an imbalance in the number of Cooper pairs (and thus in the electric charge) between the superconducting islands. In the presence of an external electromagnetic field of the microwave cavity, 
the tunneling amplitude of the Cooper pair from the left island to the right one is multiplied by the Wilson line   $\exp\{i 2e  \int dx A_1\}$, and the tunneling amplitude from the right to left island is multiplied by $\exp\{- i 2e  \int dx A_1\}$; the factor $2$ in the phase is due to the $2e$ charge of the Cooper pair. 

The difference in the tunneling amplitudes can create an imbalance between the number of Cooper pairs in the left $N_L$ and right $N_R$ islands. In general, the state of this transmon, or charge, qubit is a quantum superposition of quantum states with different imbalance $N_R - N_L$. Both in the case of the transmon qubit and the ``chiral qubit" based on the chiral anomaly and described in section \ref{sec:qubit}, this imbalance (and thus the state of the qubit) is controlled by the expectation value of the Wilson line operator. 
The charge-imbalanced quantum state of the transmon qubit is thus analogous to the chirally imbalanced state discussed above; we will now make this analogy explicit.

\section{The chiral qubit}
\label{sec:qubit}

Consider a thin ring of radius $R \sim 1\ \mu$m and thickness $d \ll R$ made of a Dirac or Weyl semimetal shown in Fig. \ref{chiral_qubit}. 
Let this ring interact with an external electromagnetic field of THz frequency. Since 1 THz corresponds to the wavelength of $0.3$ mm,  
the wavelength of electromagnetic field $\lambda$ is much larger than the thickness of the ring $d \ll \lambda$, and we can model the system by the lagrangian (\ref{lagr}), (\ref{lagr_ferm}) describing the interaction of $(1+1)$ dimensional fermions with the $(3+1)$ dimensional gauge field. In this setup, the electric field entering the anomaly equation (\ref{anom_eq}) can be seen as resulting from the change in time of the loop integral $\oint A_1 dx$ that, by Stokes theorem, is equal to magnetic flux threading the ring. Using, as before, the Coulomb gauge and the polar coordinates, we can relate the magnetic flux $\Phi$ through the ring to the gauge potential by choosing $A_\theta = \Phi/(2\pi R)$. 
\vskip0.3cm

The Dirac Hamiltonian corresponding to  (\ref{lagr_ferm}) then takes the form 
\be\label{ham}
\hat{H} = \hbar \omega \left(-i \partial_\theta + \frac{\Phi}{\Phi_0}\right) \hat{\sigma}_z ,
\ee
where $\omega = v_F/R$ and $\Phi_0 \equiv h/e$ is the flux quantum, and we have restored the Planck constant $h$. The Hamiltonian of this type has been previously considered e.g. for the graphene rings \cite{recher2007aharonov, rozhkov2011electronic, sticlet2013persistent}.

As discussed above in section \ref{sec:anomaly}, the two eigenstates of $\hat{\sigma}_z$ in (\ref{ham}) correspond to the right- and left-handed chiral fermions. In the ring setup, they are the clockwise and counter-clockwise rotating fermions, see Fig. 2. The corresponding stationary eigenstates of the Hamiltonian (\ref{ham}) are determined by periodicity in the angle $\theta$:
\be\label{stat}
\Psi_{R,L}(\theta, t) = \exp{(-i E_n^{R,L} t + i n \theta)}\ \psi_{R,L},
\ee
where, as before, $\psi_{R,L}$ are the eigenstates of  $\sigma_z$: 
\be\label{eigen_z}
\sigma_z \psi_{R,L} = \pm \psi_{R,L},
\ee
 with 
$+$ ($-$) corresponding to right (left)-handed fermions. 

Note that (\ref{eigen_z}) does {\it not} imply that the right- and left- handed fermions have their spins polarized in the direction orthogonal to the plane of the ring as it is sometimes assumed. Indeed, there is no spin in one spatial dimension, and since in this case $\gamma_5 = \sigma_z$, the eigenstates of $\sigma_z$ simply correspond to clockwise (R) and counter-clockwise (L) rotating fermions.  

Substituting (\ref{stat}) into (\ref{ham}), we find the eigenvalues
\be\label{eigenstate}
E_n^{R,L} = \pm \hbar \omega \left(n + \frac{\Phi}{\Phi_0}\right);\ n \in \mathbb{Z},
\ee
corresponding to the positive energy particles and negative energy "antiparticles"; 
the sign of the energy is fixed by chirality. 
\vskip0.3cm

The defining features of the chiral anomaly are all present in (\ref{eigenstate}):
\begin{itemize}
\item{there is an infinite tower of eigenstates with energies extending down to $E \to - \infty$ -- this is the ``Dirac sea" of relativistic quantum field theory;}
\item{the energies of all states in this infinite tower respond to an external magnetic flux, so it is impossible to describe the Dirac sea response by introducing an ultraviolet cutoff;}
\item{Right- and left-handed fermions respond differently to an external magnetic flux -- the energies of the right-handed fermions increase with $\Phi$, whereas the energies of the left-handed fermions decrease. As a result, an external magnetic flux generates a collective current in the Dirac sea \cite{gribov2001anomalies}.}
\end{itemize}

The external flux $\Phi$ in general breaks the degeneracy between the energies 
of the right- and left-handed fermions. However, if $\Phi/\Phi_0$ is an integer or a half-integer, due to the symmetry under time reversal, 
the energy spectra of the right- and left-handed fermions are the same, and so 
each energy level is doubly degenerate due to the presence of both right- and left-handed fermions, in accord with the Kramers theorem. 

Nevertheless, if we look not at the entire spectrum but at a specific energy level, its energy does change when $\Phi/\Phi_0$ changes, even if this is a change by an integer number. As a result, as follows from (\ref{anom_eq}), a time-dependent magnetic flux through the ring $\Phi = \oint A_1 dx = R \oint A_\theta d\theta$ results in the change in the difference between the occupation numbers of right- and left-handed fermions.
\vskip0.3cm

Each of the states within the spectrum given by (\ref{eigenstate}) corresponds to a circulating charged fermion, and thus supports a corresponding quantum ``persistent" current given by
\be\label{current}
J_n^{R,L} = - \frac{\partial E_n^{R,L}}{\partial \Phi} = \mp e \frac{\hbar \omega}{2\pi},
\ee  
where the upper (lower) sign corresponds to the right-(left-)handed fermion. Since the current (\ref{current}) is the same for all quantum states $n$, the total current has to be evaluated as the sum over all occupied quantum states.

To compute the resulting spatial component $J$ of the electric (vector) current $J^\mu = J^\mu_R + J^\mu_L$, we need to perform the sum over all occupied states of left- and right-handed fermions:
\be\label{totcur}
J = J_R + J_L = e \frac{\hbar \omega}{2\pi} \left(\sum_{n=-\infty}^{N_L} 1 -  \sum_{m=-\infty}^{N_R} 1 \right), 
\ee
where $N_L$ ($N_R$) is the maximal value of the quantum number $n$ for the occupied left (right)-handed fermions, and $1$ is the occupation number for the n-th fermion mode. It is clear that the individual terms in (\ref{totcur}) diverge; however their difference is finite. The Fermi energies of left- and right-handed fermions are given by $E_F^L = E_n^L(n = N_L)$ and $E_F^R = E_n^R(n = N_R)$; introducing the {\it chiral chemical potential} $\mu_5 =  (E_F^R - E_F^L)/2$, we can re-write (\ref{totcur}) as
\be\label{cme2}
J = - e\ \frac{\mu_5}{\pi}.
\ee
This is the formula for the chiral magnetic current \cite{Kharzeev2008} in $(1+1)$ dimensions \cite{alekseev1998universality,kharzeev2011chiral}; see \cite{Kharzeev2014} for discussion. 
\vskip0.3cm

In $(3+1)$ dimensions, the chiral magnetic effect requires an external magnetic field $\vec{B}$; in this case the current (\ref{cme2}) flows in the direction of $\vec{B}$. To obtain the current density in $(3+1)$ dimensions, (\ref{cme2}) has to be multiplied by the density of fermion states in the plane transverse to $\vec{B}$ given by $eB/(2\pi)$; this yields
\be\label{cme4}
\vec{J} = - \frac{e^2 \mu_5}{2 \pi^2}\ \vec{B},
\ee
in agreement with \cite{Kharzeev2008}. 

At finite temperature, to evaluate the current one needs to differentiate not the energy but the thermodynamic potential; however the results for the current (\ref{cme2}), (\ref{cme4}) do not change and are independent of temperature \cite{Kharzeev2008} -- this is a consequence of the chiral anomaly that relates the chirality of fermion modes to the topology of gauge field. This means that the CME is robust with respect to all perturbations that do not flip the chirality of fermions.
\vskip0.3cm

We have thus established that the current circulating in the ring (or in a three-dimensional crystal) is determined by the difference in the Fermi energies of right- and left-handed fermions -- the chiral chemical potential. This difference can be controlled through the chiral anomaly by a time-dependent  magnetic flux through the ring (in $1+1$ dimensions) or by circularly polarized photons (in $3+1$ dimensions) \cite{kaushik2019chiral}. 

Indeed, 
consider a right-handed eigenstate of the Hamiltonian (\ref{ham}). If the magnetic flux is constant, this eigenstate is stationary, and corresponds to a persistent clockwise quantum current (\ref{current}) circulating along the ring. Let us now introduce a time-dependent external magnetic flux. This time-dependent flux will induce an electric field along the ring, 
\be\label{flux}
\hspace{-0.5cm}\int dx E_1 = - \frac{d}{dt}\oint dx A_1 = - R \frac{d}{dt}\oint d\theta A_\theta = - \frac{d \Phi}{dt}, 
\ee
which is nothing but the Faraday law. 
Combining (\ref{flux}) with the anomaly equation (\ref{anom_eq}), we get 
\be
\frac{d(N_R-N_L)}{dt} = \frac{e}{\pi}\ \frac{d \Phi}{dt}, 
\ee
which shows that a time-dependent magnetic flux through the ring will change the chirality of fermions circulating around the ring. In the course of an adiabatic change in magnetic flux by some value $\Delta \Phi$, the change $\Delta(N_R-N_L)$ in the difference in the occupation number of right- and left-handed fermions will be given by
\be\label{change}
\Delta(N_R-N_L) =  \frac{e}{\pi}\ \Delta \Phi.
\ee
Because the gauge field responsible for magnetic flux in our case is external and ``lives" in $(3+1)$ dimensions, there is no reason for the r.h.s. of (\ref{change}) to be integer. On the other hand, the l.h.s. of (\ref{change}) is the difference in the number of right- and left-handed fermions that naively is expected to be integer. 
\vskip0.3cm

The solution to this apparent contradiction is the following: the anomaly equation
(\ref{anom_e}) has to be understood as an operator relation since the axial current $J_\mu^A$ is a composite operator made of the operators of fermion fields. Therefore, the axial charge $Q_A =  N_R - N_L$ is an expectation value of the composite operator $J_0^A = \int dx (\psi^\dagger_R \psi_R - \psi^\dagger_L \psi_L)$ taken over the entire Dirac sea that in general is a superposition of the right- and left-handed fermions moving in the background of electromagnetic field, so that (\ref{change}) can be satisfied at any $\Delta \Phi$. Note that the expectation values of $\int dx \psi^\dagger_R \psi_R$ and $\int dx \psi^\dagger_L \psi_L$ both diverge due to an infinite number of states in Dirac sea, but their difference is finite and given by (\ref{change}).

\vskip0.3cm
Let us evaluate the energy $U(\Phi)$ of the system by summing over the energies (\ref{eigenstate}) of the individual states. The corresponding sum is of course divergent but can be formally performed using the heat kernel regularization method, see \cite{manton1985schwinger}. Omitting the infinite term and a constant $-1/12$, the result is
\be
U(\Phi) = \hbar \omega \left(\frac{\Phi}{\Phi_0} - \frac{1}{2}\right)^2, \frac{\Phi}{\Phi_0} \in (0,1) .
\ee
This energy is minimal at a non-zero value of magnetic flux $\Phi = \Phi_0/2$, for which the Wilson line is $W[A] \equiv \exp\{i e  \int dx A_1\} = -1$. 

For arbitrary value of the flux, the periodicity in $\Phi$ leads to 
\be\label{eff_potential}
U(\Phi) = \hbar \omega \left(\frac{\Phi}{\Phi_0} - \frac{1}{2} - \left\lfloor{\frac{\Phi}{\Phi_0} - \frac{1}{2}}\right\rfloor \right)^2.
\ee
This periodic ``washboard" potential emerges also in the study of the vacuum energy of non-Abelian gauge theories \cite{witten1998theta,kharzeev1998possibility,kharzeev2001aspects}.
\vskip0.3cm
Let us now note that the potential (\ref{eff_potential}) has the form that is identical to the potential of a superconducting ring, see e.g. \cite{friedman2000quantum}. Moreover, adding to (\ref{eff_potential}) the term $- \Delta (\exp(i \Phi/\Phi_0) + \exp(- i \Phi/\Phi_0)) = - 2 \Delta \cos (\Phi/\Phi_0)$ \cite{kharzeev1998possibility,kharzeev2001aspects} describing the tunneling transitions between the right- and left-handed fermions with probability determined by parameter $\Delta$ (e.g., the mass gap if the fermions have a finite mass), 
transforms  (\ref{eff_potential}) into the form identical to the potential of a SQUID or two Cooper boxes separated by a Josephson junction used as a superconducting qubit \cite{friedman2000quantum,girvin2011circuit}:
\be\label{pot_tot}
U_{tot}(\Phi) = U_0 \left[\left(\frac{\Phi}{\Phi_0} - \frac{1}{2}\right)^2 - \beta \cos\left(\frac{\Phi}{\Phi_0}\right)\right].
\ee
For a SQUID, the first term describes the magnetic energy of the loop, and the second term -- the Josephson coupling energy of the junction. The parameter $U_0$ for the SQUID is inversely proportional to the inductance of the loop, and $\beta$ is proportional to the product of inductance and the critical current. The ground states of the potential (\ref{pot_tot}), (\ref{eff_potential}) are the states with different flux $\Phi$ that contain the right- or left-handed fermions. The symmetric and antisymmetric combinations of the right- and left-handed states 
\be\label{qubit}
|0\rangle = \frac{1}{\sqrt{2}} \left( |R\rangle + |L\rangle \right), |1\rangle = \frac{1}{\sqrt{2}} \left( |R\rangle - |L\rangle \right).
\ee
will have different energies as a consequence of tunneling between the right- and left-handed states.
\vskip0.3cm

This analogy with the Hamiltonian of of the superconducting qubit allows us to propose the ring supporting chiral fermions as the ``chiral qubit". The qubit states (\ref{qubit}) are superpositions of the left- and right-handed chiral fermions, and are controlled by an external magnetic flux. For a bulk crystal, this control can be achieved using circularly polarized light \cite{kaushik2019chiral}. The Hamiltonian of a thin ring considered above can also be realized for chiral edge states of the Spin Hall insulators. 

\section{Conclusions and Outlook}
\label{sec:outlook}

The main advantages of the proposed chiral qubit as compared to conventional superconducting devices are the following: i) it can operate at much higher, and potentially room, temperature; ii) the operation frequency is in the $\sim$ 10 THz range, which is about $10^4$ higher than for the superconducting qubits. At the same time, for the thin ring architecture, the Hamiltonian of the chiral qubit is similar to that of the superconducting qubit. This means that quantum gates can be implemented in a traditional way, and the algorithms developed for superconducting quantum processors will apply. 
\vskip0.3cm

The physical implementation of quantum gates in the chiral qubit will require the use of THz light. Recent experimental observations \cite{Ma2017,gao2019coherent} of photocurrents in Weyl semimetals driven by THz lasers show that this is possible. The observed \cite{gao2019coherent} near field generated by the chiral photocurrent can be utilized for entangling the chiral qubits. 
\vskip0.3cm

The main challenge in implementing the chiral qubit is a relatively short coherence time of chiral states -- the characteristic chirality flipping rate in the presently available chiral materials as inferred from the magnetotransport measurements \cite{Li2016,Xiong2015,Huang2015} is in the 1 - 100 GHz frequency range. Since the quantum gates in chiral qubits are expected to operate at $\sim$ 10 THz frequency, the ratio of the coherence time to gate time may reach $10^4$. Nevertheless it is highly desirable to identify materials with the minimal rate of chirality-flipping transitions. In addition to Dirac and Weyl semimetals, these materials may also include Quantum Spin Hall insulators in which the dynamics of chiral edge states can be described by the thin ring Hamiltonian considered in section \ref{sec:qubit}.
\vskip0.3cm

In spite of the significant challenges in practical realization of the quantum processor based on the chiral qubits, we believe that the promise of a THz frequency, room temperature quantum computer makes it worthwhile to further investigate the proposal outlined above.

\bibliographystyle{my-refs}

\section*{Acknowledgements}
We thank \mbox{S. Bravyi}, J. Chow, J. Gambetta, and \mbox{M. Steffen} for stimulating discussions and hospitality during our visit to IBM-Q. We are grateful also to 
\mbox{S. Kaushik,} M. Liu, and \mbox{E. Philip} for useful discussions. 

The work of D.K. was supported by the U.S. Department of Energy, Office of Nuclear Physics, under contracts DE-FG-88ER40388 and DE-AC02-98CH10886, and by the Office of Basic Energy Science under contract DE-SC-0017662.
The work of Q.L. was supported by the US Department of Energy, Office of Basic Energy Science, Materials Sciences and Engineering Division, under contract DE-SC0012704.

\bibliography{dirac}
\end{document}